\documentclass[fleqn,twoside]{article}
\usepackage{espcrc2}

\usepackage{graphicx}
\usepackage[figuresright]{rotating}

\usepackage{amsfonts}
\usepackage{latexsym}
\usepackage{tabularx}
\usepackage{array}
\usepackage{graphics}
\usepackage{epsfig}
\usepackage{amssymb}

\newcommand{\AmS}{{\protect\the\textfont2
  A\kern-.1667em\lower.5ex\hbox{M}\kern-.125emS}}

\newcommand{\bea}{\begin{eqnarray}}
\newcommand{\eea}{\end{eqnarray}}
\newcommand{\beq}{\begin{equation}}
\newcommand{\eeq}{\end{equation}}
\newcommand{\beqa}{\begin{eqnarray*}}
\newcommand{\eeqa}{\end{eqnarray*}}

\newcommand{\gev}{{\rm GeV}}

\newcommand{\gbb}{g_{_{B^*B\pi}}}
\newcommand{\gdd}{g_{_{D^*D\pi}}}
\newcommand{\ghh}{g_{_{H^*H\pi}}}
\newcommand{\ghat}{\widehat g}

\newcommand{\mhs}{m_{H^*}}
\newcommand{\mhm}{{\overline{m}}_{H}}
\newcommand{\mdm}{{\overline{m}}_{D}}

\title{A lattice estimate of the $\gdd$ coupling}

\author{A.~Abada\address[LPT]{Laboratoire de Physique
		Th\'eorique$^\dagger$ \\
		Universit\'e Paris-Sud, B\^at. 210, 91405 Orsay Cedex,
		France.},  
	D.~Be\'cirevi\'c\address{Dipartimento di Fisica, Universit\`a di Roma ``La Sapienza",\\ 
Piazzale Aldo Moro 2, I-00185 Rome, Italy.},
        Ph.~Boucaud\addressmark[LPT],
	G.~Herdoiza\addressmark[LPT],
	J.P.~Leroy\addressmark[LPT],
	A.~Le~Yaouanc\addressmark[LPT],
	O.~P\`ene\addressmark[LPT]
	~and~
	J.~Rodr\'{\i}guez--Quintero\address{Dpto. de F\'{\i}sica Aplicada e 
		Ingenier\'{\i}a el\'ectrica \\
		E.P.S. La R\'abida, Universidad de Huelva, 
		21819 Palos de la fra., Spain.}
	\thanks{Presented by Gregorio Herdoiza.
		\hfill \break 
		$^\dagger$Unit\'e Mixte de Recherche du CNRS - UMR 8627.}
        }

\begin{document}

\begin{abstract}
\noindent

We present the results of the first direct determination of the $\gdd$ coupling using lattice QCD. From
our simulations in the quenched approximation,  we obtain $\gdd = 18.8 \pm 2.3^{+1.1}_{-2.0}$ and
$\ghat = 0.67 \pm 0.08^{+0.04}_{-0.06}$. It is in agreement with a recent experimental result
from CLEO. 

\end{abstract}

\maketitle


\section*{Introduction}
\label{intro}

The strong coupling constant $\ghh$ of heavy-light mesons\footnotemark[1]\footnotetext[1]{We use in a
generic way the letter $H$ for the $D$ or the $B$ mesons.} and a soft pion is one of the essential
parameters entering the lagrangian which englobes both the chiral and the heavy quark
symmetry~\cite{Casalbuoni:1996pg}. Its precise value is useful in evaluating the nearest pole
contribution to the shape of the $B \to \pi$ and $D \to \pi$ semileptonic decay form factor, since the
residua are directly proportional to $\gbb$ and $\gdd$, respectively. As $\gbb$ is not measurable
experimentally due to phase space, its theoretical prediction is useful to restrict the
uncertainties on quantities like the $\vert V_{ub}\vert$ CKM matrix element.

The coupling $\ghh$ is related to the coupling constant $\ghat$ appearing in the chiral theory
for heavy mesons, 
\bea \label{Def}
\ghh\ = \  { 2 \sqrt{\mhs m_H} \over  f_{\pi} } \ \ghat \,. 
\eea
\noindent The coupling $\ghat$ is a constant up to small $\mathcal{O}\left( {1 / m_H } \right)$
corrections in the heavy mass. A large number of predictions for $\ghat$, using several methods, can be
found in the literature (see~\cite{Becirevic:1999fr} for a list of results). A surprising feature in
these determinations is the discrepancy between QCD sum rules which find typically low values, $\ghat \sim 0.3$,
and quark models which naturally accommodate larger values\footnotemark[2]\footnotetext[2]{The
Adler-Weisberger sum rule sets the bound $\ghat<1$.}. Recently, the CLEO collaboration measured the
coupling $\gdd$ and deduced, from eq.(\ref{Def}), a value for $\ghat$~(see ref~\cite{CLEO}):
\bea \label{Result_CLEO}
\gdd &=& 17.9\pm 0.3\pm 1.9\,, ~~~~ {\rm i.e.} \nonumber \\ \ghat &=& 0.59 \pm 0.07 \,.
\eea
An exploratory lattice computation of $\ghat$ has been  performed in ref.~\cite{UKQCD}. That
computation has been made in the static  limit of HQET. To be able to directly confront the lattice  to
the experimental result, we decided to make the lattice QCD  simulation by using fully relativistic
quarks in the region of the charm quark mass. Notice that, contrary to the $b$-quark, the charmed quarks
are directly accessible from the currently available lattices. In other words, to confront to the
experiment, no extrapolation in the heavy quark mass is needed.

\section{Theoretical basis}
\label{th}

To determine $\ghh$, we compute the matrix element of the axial vector current $A_\mu = \overline{q}
\gamma_\mu \gamma_5 q$, between the vector ($H^{\ast}$) and pseudoscalar ($H$) heavy-light mesons. This
matrix element is parametrized by the three form factors $A_{0,1,2}(q^2)$, where $q$ is the momentum
transfer between $H^{\ast}$ and $H$.

The coupling $\ghh$ is related to the form factors through the following expression (see ref.[5] for
details):
\bea \label{gA12}
	\ghh = \frac{1}{f_\pi}~\times &[~& \left( m_{H^{\ast}}+m_{H} \right) A_1(0) \cr
	  &~~+& \left(m_{H^{\ast}}-m_{H}\right) A_2(0)~] \, .
\eea
The first term ($\propto A_1(0)$) is dominant since the heavy masses $m_{H^{\ast}}$ and $m_{H}$  are
almost degenerate and, in the kinematical region where $q^2 \sim 0$, $A_1$ and $A_2$ are of the same
order of magnitude. Our simulation has confirmed that the subdominant term ($\propto A_2(0)$) is indeed
small ($\lesssim 5\%$). In studying $A_1(q^2)$, the point $q^2 \sim 0$ is reached already  at $\vec q=\vec
0$. On the lattice, it is known that the signals are better when no external momenta are given to the
interacting hadrons. All these conditions make the determination of $\gdd$ rather clean.  

\section{Lattice simulation}
\label{latt}

The main results are obtained from a simulation on a $24^3\times 64$ lattice, at $\beta =6.2$, using
$100$ configurations produced in the quenched approximation with the non-perturbative ${\cal O}(a)$
improvement for Wilson fermions.

We have used six values of the light quark masses, corresponding to a pseudoscalar meson mass $\in
[0.59, 0.83 ]~\gev$, and three values of the heavy quark masses, corresponding to $\mhm \in [ 1.83,2.32
]~\gev$, where, $\mhm = {(3\mhs + m_H)}/{4}$, is the spin averaged mass of the heavy meson. Note that this
interval includes the value $\mdm = 1.97~\gev$, so that no heavy mass extrapolation is required to reach
the $\bar D$-meson.

We inserted to the current $A_\mu$ the momentum  $\vec{q}=\{(0,0,0);(1,0,0)\}$ (in ${2\pi}/{L}$
units), and we have checked that the case ${\vec q} = {\vec 0}$ corresponds indeed to small values of
$q^2$ (we have $q^2\lesssim 0.01~\gev^2$).

\begin{figure}[tb]
\vspace*{-7.5mm}
\includegraphics[width=7.4cm,height=4.0cm]{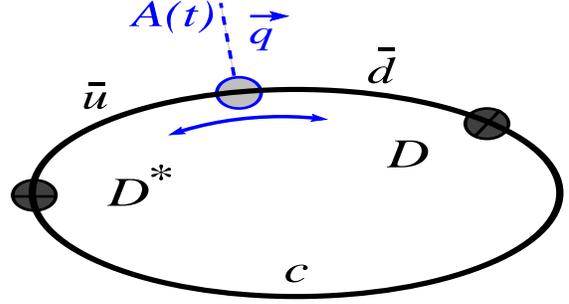}
\vspace*{-9mm}
\caption{\label{fig1}\sl \small Three--point Green function computed on the lattice: $\langle D |
A^{\mu} | D^{\ast} \rangle$.}
\label{fig_3pt}
\vspace*{-7.5mm}
\end{figure}

The matrix elements that we compute on the lattice are illustrated in fig.~\ref{fig1}. We extract
the form factors $A_1$ and $A_2$ by using the standard analysis of the two-point and three-point Green
functions on the lattice.

In order to determine $\gdd$, we first have to {\it extra}polate our simulated light quark masses down
to the $u/d$-quark (corresponding to the physical pion) and then {\it inter}polate the heavy meson
masses to the physical $D$ meson.

\begin{figure}[tb]
\vspace*{-1.0mm}
\includegraphics[width=7.4cm,height=5.0cm]{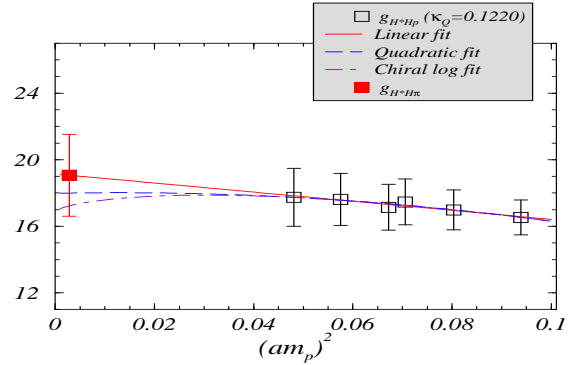}
\vspace*{-9mm}
\caption{\label{fig2}\sl \small Illustration of the chiral extrapolation of $\ghh$ for a given value of
the heavy quark Wilson hopping parameter $\kappa_Q$. The lower subscript ``p''  labels the light
pseudoscalar meson.}
\label{fig_3pt}
\vspace*{-7.5mm}
\end{figure}

The chiral extrapolation has been performed according to several forms (linear, quadratic, ``chiral log'';
see fig.~\ref{fig2}). In our region of light quark masses, we are not able to isolate non-linearities.
Therefore, we shall use non-linear fits to estimate the systematic uncertainties.

The heavy mass interpolation function is guided by the heavy quark symmetry, $\ghat = a + b / \mhm$.
The value of the parameter $a$ corresponds to the value of $\ghat$ in the static limit, which we note
$\ghat_\infty$.

As a comment, we remark that the computation of the form factors $A_{1,2}$, and therefore $\gdd$
(eq.(\ref{gA12})), do not depend on the improvement of the bare axial current $A_\mu$ (see
ref.~\cite{Abada:2002xe}).

\section{Results}
\label{res}

We present in table~\ref{tab1}, the values of $\ghh$ and $\ghat$ for each heavy-light
meson mass $\mhm$. All these quantities are already linearly extrapolated to the $u/d$-quark mass.

\begin{table}[tb]
\begin{center}
\begin{tabular}{@{}|c|c c c|}
\hline
{$\mhm~[\gev~]$} & $1.83(9)$ & $2.08(10)$ & $2.32(11)$ \\ \hline \hline
\phantom{\large{l}} \raisebox{0.2cm}{\phantom{\large{j}}}
{$\ghh$} & $17.7 \pm 2.2$ & $20.1 \pm 2.7$ & $22.6 \pm 3.3$  \\
\phantom{\large{l}} \raisebox{0.3cm}{\phantom{\large{j}}}
{$\ghat$} & $0.669(72)$ & $0.668(79)$ & $0.673(88)$  \\
\phantom{\large{l}} \raisebox{0.1cm}{\phantom{\large{j}}} 
~&~&~& \\
\hline
\end{tabular}
\vspace*{2.0mm}
\caption{\label{tab1} \small{\sl For each heavy-light meson mass $\mhm$, we show the values of $\ghh$ and of the corresponding
$\ghat$.}}\
\vspace*{-10.0mm}
\end{center}
\end{table}

\begin{figure}[tb]
\vspace*{-2mm}
\includegraphics[width=7.4cm,height=4.5cm]{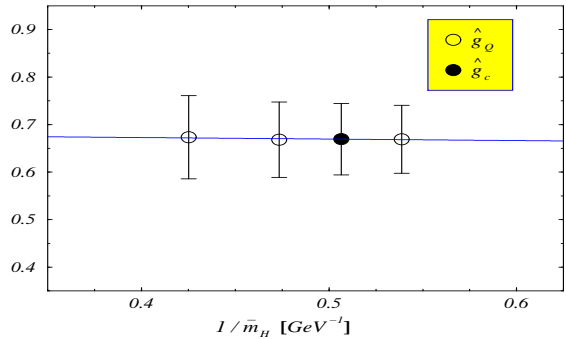}
\vspace*{-10mm}
\caption{\label{fig3}\sl \small Interpolation of $\ghat$ to the  $\bar D$-meson (filled circle).}
\label{fig_3pt}
\vspace*{-7.5mm}
\end{figure}

The heavy quark interpolation of $\ghat$ to the $\bar D$-meson mass is shown in fig.~\ref{fig3}. Our
results indicate that the slope in $1/m_H$  for the coupling $\ghat$ is small. Assuming that the linear
dependence in $1/m_H$ holds all the way to the static limit, $1/m_H \to 0$, we obtain that $\widehat
g_\infty$ is not more than $10$~\% larger than $\widehat g$ at the level of the charm quark mass.

Our final results, at $\beta = 6.2$, are:
\bea \label{Result}
\gdd &=& 18.8 \pm 2.3^{+1.1}_{-2.0}\,, ~~~~ {\rm i.e.} \nonumber \\ \ghat &=& 0.67 \pm 0.08^{+0.04}_{-0.06}\,. 
\eea

\section{Systematic uncertainties}
\label{syst}

In order to study the ${\cal O}(a)$ effects, we have  performed in addition to the simulation at $\beta
= 6.2$, another one at $\beta =6.0$.  The difference between the two results is smaller than one standard
deviation. With two values of the lattice spacing we cannot attempt a continuum limit extrapolation,
but, as it is observed for other similar quantities, we can hope that the results at $\beta = 6.2$ are
close to those in the continuum limit. Further computations at larger $\beta$ would certainly improve
the precision in the determination of $\ghat$.

Finite volume effects were studied by performing simulations at $\beta =6.0$ with two different volumes,
$16^3\times 64$ and  $24^3\times 64$. Within our statistics, we do not see any evidence for the presence
of finite lattice volume effects. From the comparison between these simulations, we add a conservative
$6~\%$ contribution to the overall systematic uncertainty.

The chiral extrapolation is the major source of uncertainty in our results. We estimate this
effect by comparing the linear fit, to the quadratic and ``chiral log'' ones ({\it
c.f.} Section~\ref{latt}).

\section*{Conclusion}
\label{ccl}

We have performed the first lattice measurement of the $\gdd$ coupling. Our results~(\ref{Result}) are
in good agreement with the experimental value~(\ref{Result_CLEO}). The discrepancies with other
theoretical predictions, like the smaller values of $\ghat$ given by the QCD sum rules, still need an
explanation. Further improvements of our results would include, studying the continuum limit and
attempting an unquenched calculation of $\ghat$.

Work supported by the European Community's Human potential programme under
HPRN-CT-2000-00145 Hadrons/LatticeQCD.


\end{document}